\documentclass[twocolumn,aps]{revtex4}
\usepackage{epsf}
\usepackage{amsmath}
\usepackage{amsfonts}
\usepackage{amssymb}

\begin{document}

\title{$z \rightarrow - z$ symmetry of spin-orbit coupling and weak localization in graphene}
\author{Edward McCann and Vladimir I. Fal'ko}
\affiliation{Department of Physics, Lancaster University, Lancaster, LA1 4YB, United
Kingdom}

\begin{abstract}
We show that the influence of spin-orbit (SO) coupling on the weak localization
effect for electrons in graphene depends on the lack or presence of $z \rightarrow - z$
symmetry in the system. While for $z \rightarrow - z$ asymmetric SO coupling, disordered
graphene should display a weak anti-localization behavior at lowest temperature,
$z \rightarrow - z$ symmetric coupling leads to an effective saturation of decoherence
time which can be partially lifted by an in-plane magnetic field, thus, tending
to restore the weak localization effect.
\end{abstract}


\maketitle


The effect of spin-orbit (SO) coupling in graphene represents an example
of a stimulating theoretical study that remains difficult to detect
experimentally. The form of intrinsic SO coupling in the
graphene band structure, suggested by Kane and Mele \cite{kanemele05},
has fuelled the theory of $Z_2$ topological insulators,
but its strength is too weak for detection by conventional
spectroscopic methods \cite{kanemele05,dhh06,min06,yao07,dhh09}.
Here, we show how the presence of SO coupling in disordered graphene could be manifested
in quantum transport measurements.
Specifically for graphene, the presence of SO coupling may not necessarily lead
to anti-localization behavior known in semiconductors and metals \cite{LarkinWAL},
and this reflects the presence or lack of $z \rightarrow -z$ symmetry
in the source of SO coupling.

In general, weak localization is very sensitive to symmetry
breaking in the electronic system or to scattering involving electron spin
since it is formed by electrons propagating along long diffusive trajectories
\cite{WL,LarkinWAL,AndoWL,mor06,guinea06,WLmono,WLreview,tik08,koz11,lara11,imura09}.
The typical behavior of electrons in metals with strong SO coupling
results in a pronounced weak antilocalization
effect manifested by positive magnetoresistance, in contrast to simple
metals and semiconductors where the weak localization magnetoresistance
caused by the interference correction to conductivity is negative.
For graphene with broken $z \rightarrow -z$ symmetry, by a substrate or deposits,
we find that, at the lowest temperatures, SO interaction leads to the
conventional weak antilocalization.
In contrast, for a $z \rightarrow -z$ symmetric system, SO coupling leads only to
a saturation in the size of the weak localization correction
rather than antilocalization, which can be taken for
a saturation of $\tau_{\varphi}(T)$ as $T \rightarrow 0$.
Then, we analyze the influence of an in-plane magnetic field on the interference correction
to conductivity for both $z \rightarrow -z$ symmetric and antisymmetric SO
coupling scenarios and find that it lifts the aforementioned saturation of $\tau_{\varphi}$.

The breaking of $z \rightarrow -z$ symmetry in pristine graphene is usually
associated with the addition of a Bychkov-Rashba term ${\hat{h}}_{BR}$ \cite{kanemele05,rashba,dhh06,min06,rasbagraphene,imura09,dhh09,cn09}
to the intrinsic SO coupling ${\hat{h}}_{KM}$ \cite{kanemele05,min06,yao07,imura09}
in the graphene Hamiltonian:
\begin{eqnarray}
&& \!\!\! \!\!\! \!\!\! \!\!\!
\!\!\! \!\!\! \!\!\! \!\!\!
\!\!\!
{\hat{H}} \!=\! v \mbox{\boldmath$\Sigma$} \mathbf{p} + {\hat{h}}_{KM}
+ {\hat{h}}_{BR} + {\hat{U}} + {\hat{V}}_{\mathrm{sym}} + {\hat{V}}_{\mathrm{asy}}
+ \mu_B \mathbf{s}\mathbf{B}_{\parallel} , \nonumber \\
{\hat{h}}_{KM} &=& \lambda \Sigma_{z} s_{z} , \qquad
{\hat{h}}_{BR} = \mu \left( \Sigma_{x} s_{y} - \Sigma_{y} s_{x} \right) . \label{h1-2}
\end{eqnarray}
The last term in ${\hat{H}}$ accounts for Zeeman splitting
due to in-plane magnetic field $\mathbf{B}_{\parallel} = {\vec{\ell}} B_{\parallel}$,
and the terms
\begin{eqnarray}
{\hat{U}} &=& u_0 \, \mathrm{\hat{I}}  + \sum_{a,l=x,y,z} \! u_{a,l} \,
\Sigma_{a} \Lambda_{l} , \label{nmdis} \\
{\hat{V}}_{\mathrm{sym}} &=& s_{z} \Big[\sum_{a=x,y,z} \! \alpha_{a,z} \Sigma_{a} +
\sum_{l=x,y,z} \! \beta_{l,z} \Lambda_{l} \Big] , \label{vz} \\
{\hat{V}}_{\mathrm{asy}} &=& \sum_{j=x,y} s_{j} \Big[ \sum_{a=x,y,z} \alpha_{a,j} \Sigma_{a}
+ \sum_{l=x,y,z} \beta_{l,j} \Lambda_{l} \Big] , \label{vp}
\end{eqnarray}
describe three types of disorder on the honeycomb lattice:
spin-independent perturbations, SO coupling with $z \rightarrow -z$ symmetric
perturbations, and $z \rightarrow -z$ asymmetric perturbations,
respectively.
Here, we use a symmetry-based approach to determine how electronic spin may be combined
with lattice and valley degrees of freedom.
We focus on the region near the Fermi level which lies in the vicinity of two
inequivalent corners of the Brillouin zone, known as valleys, with
wave vectors $\mathbf{K}_{\pm} = \pm ( 4\pi / 3a , 0 )$ where $a$ is the lattice constant,
and the momentum measured from the
center of a valley is $\mathbf{p} = \hbar\mathbf{k} - \hbar\mathbf{K}_{\pm}$.
The Hamiltonian (\ref{h1-2}) operates in a space of eight-component Bloch functions
$\Phi
=$[$\phi_{\mathbf{K}_{+},A,\uparrow}$,
$\phi_{\mathbf{K}_{+},B,\uparrow}$,
$\phi_{\mathbf{K}_{-},B,\uparrow}$,
$\phi_{\mathbf{K}_{-},A,\uparrow}$,
$\phi_{\mathbf{K}_{+},A,\downarrow}$,
$\phi_{\mathbf{K}_{+},B,\downarrow}$,
$\phi_{\mathbf{K}_{-},B,\downarrow}$,
$\phi_{\mathbf{K}_{-},A,\downarrow}$
consisting of two valleys
$\mathbf{K}_{+}$/$\mathbf{K}_{-}$, two lattice sites $A$/$B$,
and two spin components $\uparrow$/$\downarrow$.
We use three sets of Pauli matrices \cite{WLmono,WLreview} to describe
spin $\vec{s}=(s_{x},s_{y},s_{z})$, sublattice `isospin'
$\vec{\Sigma}=(\Sigma_{x},\Sigma_{y},\Sigma_{z})$
and valley `pseudospin'
$\vec{\Lambda}=(\Lambda_{x},\Lambda_{y},\Lambda_{z})$ \cite{matrices}.
The matrices $\vec{s},\vec{\Sigma},\vec{\Lambda}$
all change sign upon time inversion
so that their products are time-inversion symmetric and
$\Sigma_{a}s_{j}$, $s_{j}\Lambda_{l}$ may
be used as a basis for a phenomenological description of static disorder
leading to SO scattering.

\begin{table}[tbp]
\caption{Irreducible representations of the planar group $C_{6v}^{\prime\prime}$ \cite{groups},
as provided by matrices $\Sigma_a$, $\Lambda_l$, and $s_j$.
Representations $A_{1}$, $A_{2}$, $B_{1}$, $B_{2}$, $E_{1}$, $E_{2}$ are part of the
point group of two-dimensional graphene $C_{6v}$, representations $E_{1}^{\prime}$,
$E_{2}^{\prime}$, $G^{\prime}$ incorporate primitive translations.}\label{tab:1}
\begin{tabular}{|c|c|c|}
  \hline
  Irr. Rep. & $z \rightarrow - z$ symmetric & $z \rightarrow - z$ asymmetric \\
\hline
  $A_{1}$ & $\hat{I}$, $\Sigma_z s_z$ & $\Sigma_x s_y - \Sigma_y s_x$ \\
  $A_{2}$ & $\Sigma_z$, $s_z$ & $\Sigma_x s_x + \Sigma_y s_y$ \\
  $B_{1}$ & $\Lambda_z$ & \\
  $B_{2}$ & $\Sigma_z\Lambda_z$, $\Lambda_zs_z$ & \\
  $E_{1}$ & $\left( \!\begin{array}{c}
  \Sigma_x \\
  \Sigma_y \\
\end{array}\!\! \right)$ \!\!, \!\!\!
$\left( \!\begin{array}{c}
  \Sigma_xs_z \\
  \Sigma_ys_z \\
\end{array}\!\! \right)$ & $\left( \!\begin{array}{c}
  s_x \\
  s_y \\
\end{array}\!\! \right)$ \!, \!\!
$\left( \!\begin{array}{c}
  \Sigma_zs_x \\
  \Sigma_zs_y \\
\end{array}\!\! \right)$ \\
  $E_{2}$ & $\left( \!\begin{array}{c}
  \Lambda_z\Sigma_x \\
  \Lambda_z\Sigma_y \\
\end{array}\!\! \right)$ &
  $\left( \!\begin{array}{c}
  \Sigma_x s_y + \Sigma_y s_x \\
  \Sigma_x s_x - \Sigma_y s_y \\
\end{array}\!\! \right)$ \!\!,\!\!
  $\left( \!\begin{array}{c}
  \Lambda_zs_x \\
  \Lambda_zs_y \\
\end{array}\!\! \right)$ \\
  $E_{1}^{\prime}$ & $\left( \!\begin{array}{c}
  \Lambda_x\Sigma_z \\
  \Lambda_y\Sigma_z \\
\end{array}\!\! \right)$ \!\!,\!\! $\left( \!\begin{array}{c}
  \Lambda_xs_z \\
  \Lambda_ys_z \\
\end{array}\!\! \right)$ & \\
  $E_{2}^{\prime}$ & $\left( \!\begin{array}{c}
  \Lambda_x \\
  \Lambda_y \\
\end{array}\!\! \right)$ & \\
  $G^{\prime}$ & $\left(
                   \begin{array}{c}
                     \Lambda_x\Sigma_x \\
                     \Lambda_x\Sigma_y \\
                     \Lambda_y\Sigma_x \\
                     \Lambda_y\Sigma_y \\
                   \end{array}
                 \right)$ & $\left(
                   \begin{array}{c}
                     \Lambda_xs_x \\
                     \Lambda_xs_y \\
                     \Lambda_ys_x \\
                     \Lambda_ys_y \\
                   \end{array}
                 \right)$ \\
  \hline
\end{tabular}
\end{table}

The results of this symmetry-based approach are summarized in Table~\ref{tab:1}
which shows how $\vec{s},\vec{\Sigma},\vec{\Lambda}$ may be combined
to form irreducible representations of the planar
group $C_{6v}^{\prime\prime}$ \cite{groups} which combines the point
group $C_{6v}$ of strictly two-dimensional graphene with primitive translations,
as appropriate for the description of two valleys $\mathbf{K}_{\pm}$.
Matrices $\vec{\Sigma}$ and $\vec{\Lambda}$ are confined to the two-dimensional
plane of graphene and their behavior under symmetry operations is
impervious to the third spatial dimension.
Thus, they are invariant under mirror reflection
in the graphene plane so that, in the absence of spin,
they only appear in the representations that are $z \rightarrow - z$ symmetric.
The presence of spin, however, introduces a pseudovector that lies
in three-dimensional space: $s_z$ is even under $z \rightarrow - z$ reflection,
but in-plane components $s_x,s_y$ are odd.
Thus, SO terms containing $s_z$ appear in $z \rightarrow - z$ symmetric
representations, terms containing $s_x,s_y$ appear in $z \rightarrow - z$ asymmetric representations.

In Table~\ref{tab:1}, $\Sigma_z s_z$ is an invariant of the point group of graphene
representing intrinsic Kane-Mele SO coupling ${\hat{h}}_{KM}$ \cite{kanemele05,min06,yao07,imura09},
and $\Sigma_x s_x - \Sigma_y s_y$ describes the Bychkov-Rashba term ${\hat{h}}_{BR}$ \cite{kanemele05,rashba,dhh06,min06,rasbagraphene,imura09,dhh09,cn09} which assumes
the existence of a transverse field $\vec{\ell}_z$ breaking $z \rightarrow -z$ symmetry.
The entries in Table~\ref{tab:1} take into account possible SO scattering mechanisms
due to defects in graphene: ${\hat{V}}_{\mathrm{sym}}$ includes terms proportional to
$s_z$, and ${\hat{V}}_{\mathrm{asy}}$ includes $s_x,s_y$.
The term ${\hat{U}}$, Eq.~(\ref{nmdis}), describes disorder decoupled from
the spin degree of freedom: $u_0(\mathbf{r})\,\mathrm{\hat{I}}$ describing
the influence of remote charges, $u_{z,z}(\mathbf{r}) \Sigma_{z}\Lambda_{z}$ describing
different on-site energies of the $A$/$B$ sublattices,
and $u_{x,z}(\mathbf{r})\Sigma_{x}\Lambda_{z}$, $u_{y,z}(\mathbf{r})\Sigma_{y}\Lambda_{z}$ accounting for fluctuations of $A$/$B$ hopping.
The other terms in ${\hat{U}}$, $u_{a,x}(\mathbf{r})\Sigma_{a}\Lambda_{x}$ and $u_{a,y}(\mathbf{r})\Sigma_{a}\Lambda_{y}$
for $a = x,y,z$, generate intervalley scattering.
We assume that different types of disorder in the Hamiltonian
are uncorrelated and $x$-$y$ isotropic:
\begin{eqnarray*}
\langle u_{a,l} \left( \mathbf{r} \right)
u_{a^{\prime},l^{\prime}} \left( \mathbf{r^{\prime}} \right) \rangle
&=& u_{a,l}^2 \delta_{a a^{\prime}} \delta_{l l^{\prime}}
\delta \left( \mathbf{r} - \mathbf{r^{\prime }} \right) , \\
\langle \alpha_{a,j} \left( \mathbf{r} \right)
\alpha_{a^{\prime},j^{\prime}} \left( \mathbf{r^{\prime}} \right) \rangle
&=& \alpha_{a,j}^2 \delta_{a a^{\prime}} \delta_{j j^{\prime}}
\delta \left( \mathbf{r} - \mathbf{r^{\prime }} \right) , \\
\langle \beta_{l,j} \left( \mathbf{r} \right)
\beta_{l^{\prime},j^{\prime}} \left( \mathbf{r^{\prime}} \right) \rangle
&=& \beta_{l,j}^2 \delta_{l l^{\prime}} \delta_{j j^{\prime}}
\delta \left( \mathbf{r} - \mathbf{r^{\prime }} \right) .
\end{eqnarray*}

\begin{table*}[tbp]
\caption{Scattering rates, due to symmetry-breaking terms in the Hamiltonian
Eq.~(\ref{h1-2}), that produce relaxation gaps $\Gamma_{j}^{l}$ in the
otherwise gapless Cooperons ${C}_{j}^{l}$ where $j$ refers to spin, $l$ to valley. The relaxation rate of the intervalley Cooperons $\tau_{\ast}^{-1}$
and the intervalley rate $\tau_{iv}^{-1}$ result from spin-independent disorder,
intrinsic $\tau_{KM}^{-1}$ and Bychkov-Rashba $\tau_{BR}^{-1}$
rates arise from coupling of spin and lattice, rates $\tau_{zv,\mathrm{e}}^{-1}$, $\tau_{zv,\mathrm{o}}^{-1}$,
$\tau_{iv,\mathrm{e}}^{-1}$, $\tau_{iv,\mathrm{o}}^{-1}$ account for coupling of valley and spin degrees of freedom.}\label{tab:2}
\begin{tabular}{|c|c|}
\hline
relaxation gaps & relaxation rates\\
\hline
{\begin{minipage}[t]{8.0cm}\begin{flushleft}
  $\Gamma_{0}^{0} = 0$ \\
  $\Gamma_{0}^{x} = \Gamma_{0}^{y} = \tau_{\ast}^{-1} + 2 \tau_{zv,\mathrm{e}}^{-1} + 4 \tau_{zv,\mathrm{o}}^{-1} + 2 \tau_{iv,\mathrm{e}}^{-1} + 4 \tau_{iv,\mathrm{o}}^{-1}$ \\
  $\Gamma_{0}^{z} = 2 \tau_{iv}^{-1} + 4 \tau_{iv,\mathrm{e}}^{-1} + 8 \tau_{iv,\mathrm{o}}^{-1}$ \\
  $\Gamma_{x}^{0} = \Gamma_{y}^{0} = \tau_{BR}^{-1} + \tau_{KM}^{-1} + 2 \tau_{zv,\mathrm{e}}^{-1} + 2 \tau_{zv,\mathrm{o}}^{-1} + 4 \tau_{iv,\mathrm{e}}^{-1} + 4 \tau_{iv,\mathrm{o}}^{-1}$ \\
  $\Gamma_{x}^{x} = \Gamma_{x}^{y} = \Gamma_{y}^{x} = \Gamma_{y}^{y} = $
\\ $\qquad \tau_{\ast}^{-1} + \tau_{BR}^{-1} + \tau_{KM}^{-1} + 2 \tau_{zv,\mathrm{o}}^{-1} + 2 \tau_{iv,\mathrm{e}}^{-1} + 4 \tau_{iv,\mathrm{o}}^{-1}$ \\
  $\Gamma_{x}^{z} = \Gamma_{y}^{z} = 2 \tau_{iv}^{-1} + \tau_{BR}^{-1} + \tau_{KM}^{-1} + 2 \tau_{zv,\mathrm{e}}^{-1} + 2 \tau_{zv,\mathrm{o}}^{-1} + 4 \tau_{iv,\mathrm{o}}^{-1}$ \\
  $\Gamma_{z}^{0} = 2 \tau_{BR}^{-1} + 4 \tau_{zv,\mathrm{o}}^{-1} + 8 \tau_{iv,\mathrm{o}}^{-1}$ \\
  $\Gamma_{z}^{x} = \Gamma_{z}^{y} = \tau_{\ast}^{-1} + 2 \tau_{BR}^{-1} + 2 \tau_{zv,\mathrm{e}}^{-1} + 2 \tau_{iv,\mathrm{e}}^{-1} + 4 \tau_{iv,\mathrm{o}}^{-1}$ \\
  $\Gamma_{z}^{z} = 2 \tau_{iv}^{-1} + 2 \tau_{BR}^{-1} + 4 \tau_{zv,\mathrm{o}}^{-1} + 4 \tau_{iv,\mathrm{e}}^{-1}$
\end{flushleft}\end{minipage}}
&
{\begin{minipage}[t]{9.5cm}\begin{flushleft}
  $\tau_{\ast}^{-1} = {\tau}_{z}^{-1} + \tau_{iv}^{-1}$ \\
  $\tau_{iv}^{-1} = \pi \gamma \left( u_{x,x}^2 + u_{x,y}^2 +
  u_{y,x}^2 + u_{y,y}^2 + u_{z,x}^2 + u_{z,y}^2 \right)/\hbar$ \\
  ${\tau}_{z}^{-1} = 2\pi \gamma \left( u_{x,z}^2 + u_{y,z}^2 + u_{z,z}^2\right)/\hbar$ \\
  $\tau_{KM}^{-1} = \lambda^2/(\epsilon_{F}^2\tau _{0})
+ 2\pi \gamma \left( \alpha_{x,z}^2 + \alpha_{y,z}^2 +
  \alpha_{z,z}^2\right)/\hbar$ \\
$\tau_{BR}^{-1} = 2\tau_{0}\mu^{2}/\hbar^{2}
+ \pi \gamma \left( \alpha_{x,x}^2 + \alpha_{x,y}^2 +
  \alpha_{y,x}^2 + \alpha_{y,y}^2 + \alpha_{z,x}^2 + \alpha_{z,y}^2 \right)/\hbar$ \\
$\tau_{zv,\mathrm{e}}^{-1} = \pi \gamma \beta_{z,z}^2 / \hbar$ \\
$\tau_{iv,\mathrm{e}}^{-1} = \pi \gamma \beta_{x,z}^2 / \hbar = \pi \gamma \beta_{y,z}^2 / \hbar$ \\
$\tau_{zv,\mathrm{o}}^{-1} = \pi \gamma \beta_{z,x}^2 / \hbar = \pi \gamma \beta_{z,y}^2 / \hbar$ \\
$\tau_{iv,\mathrm{o}}^{-1} = \pi \gamma \beta_{x,x}^2 / \hbar = \pi \gamma \beta_{x,y}^2 / \hbar = \pi \gamma \beta_{y,x}^2 / \hbar = \pi \gamma \beta_{y,y}^2 / \hbar$ \\
$\gamma = p_F/(2 \pi \hbar^2 v)$
\end{flushleft}\end{minipage}} \\
\hline
\end{tabular}
\end{table*}

In the following study, we employ the
standard diagrammatic technique for disordered systems \cite{WL,LarkinWAL}
to calculate the weak localization correction $\delta \sigma$ to the conductivity.
We assume that the Dirac-like Hamiltonian $v \mbox{\boldmath$\Sigma$} . \mathbf{p}$
dominates the electronic behavior and that diagonal disorder,
$\mathrm{\hat{I}}u_0(\mathbf{r})$ in Eq.~(\ref{nmdis}),
determines the elastic scattering rate,
$\tau^{-1} \approx \tau_0^{-1} = \pi \gamma u^2 / \hbar$,
where $\gamma = p_F/(2 \pi \hbar^2 v)$ is the density of states
per spin, per valley \cite{WLmono,WLreview}.
The current operator corresponding to the Dirac-like Hamiltonian
is momentum independent so that the current vertex entering the
Drude conductivity is renormalized by vertex corrections.
Then, the Drude conductivity is equal to
$\sigma = 4e^2 \gamma D$ where the diffusion coefficient
is $D = v^2 \tau_{tr}/2$ and the transport time is twice the
scattering time, $\tau_{tr} = 2 \tau_{0}$ \cite{AndoWL}.

The weak localization correction $\delta \sigma$
may be written in terms of
disorder-averaged two-particle correlation functions known as
Cooperon propagators $C_{s}^{l}$ where index $l$ refers to
pseudospin (related to $\vec{\Lambda}$ describing valley
degrees of freedom), and $s$ refers to spin (related to $\vec{s}$).
All the Cooperons that we consider are singlets with respect to
sublattice isospin $\vec{\Sigma}$
because all isospin-triplet modes have relaxation gaps $\sim 1/\tau_0$ \cite{WLmono,WLreview}.
Then, $\delta \sigma$
may be written in terms of a summation with respect to
sixteen Cooperons consisting of combinations of spin and pseudospin
singlet and triplets:
\begin{gather}
\delta \sigma =\frac{e^{2}D}{\pi \hbar}
\sum_{s,l=0,x,y,z} c_s c_l \, {C}_{s}^{l} \left( \mathbf{r^{\prime} = \mathbf{r}} \right) \, , \label{wlc1}\\
\left[ \! D \! \left( \! i\mathbf{\nabla} + \frac{2e\mathbf{A}}{c\hbar} \! \right)^{2} \!
\!\!\! + \Gamma_{s}^{l} + \tau_{\varphi }^{-1} - i\omega \right]
\!\! C_{s}^{l} \! \left( \mathbf{r},\mathbf{r^{\prime }} \right) =
\delta \! \left( \mathbf{r} - \mathbf{r^{\prime }} \right) . \nonumber
\end{gather}
Here, the factors $c_0 = 1$, $c_x = c_y = c_z = -1$ take into
account the fact that singlet and triplet Cooperons (of both
spin and pseudospin) appear with opposite signs,
and $\mathbf{A}$ is the vector potential of homogeneous
external magnetic field, $\mathbf{B} = \mathrm{rot} \mathbf{A}$
($B_z = \partial_x A_y - \partial_y A_x$).

Inelastic dephasing is taken into account in Eq.~(\ref{wlc1}) by $\tau_{\varphi}^{-1}$
and, in general, symmetry-breaking perturbations [such as those
contained in the Hamiltonian Eq.~(\ref{h1-2})],
contribute relaxation gaps $\Gamma_{s}^{l}$
to the otherwise gapless Cooperons ${C}_{s}^{l}$,
as quantified in terms of relaxation rates
summarized in Table~\ref{tab:2}.
Then, the zero-field temperature-dependent correction, $\delta \rho \left( 0 \right)$, to
the sheet resistance, where
$\delta \rho \left( 0 \right)/\rho^2 \equiv - \delta \sigma$, may be written as
\begin{eqnarray}
\delta \rho \left( 0 \right) &=& - \frac{e^2\rho^2 D}{\pi \hbar}
\!\!\! \sum_{s,l=0,x,y,z} \!\!\! c_s c_l \! \int_{0}^{(D\tau)^{-1/2}}
\!\!\! \frac{q\, dq}{2\pi} C_{s}^{l} (q) , \label{f1} \\
&=& - \frac{e^2\rho^2}{2\pi h}
\sum_{s,l=0,x,y,z} c_s c_l \,
\ln \left( \frac{\tau^{-1}}{\tau_{\varphi}^{-1} + \Gamma_{s}^{l}} \right) ,
\label{monr0}
\end{eqnarray}
and the magnetoresistance,
$\Delta \rho (B_z) = \delta \rho \left( B_z \right) - \delta \rho \left( 0 \right)$, as
\begin{eqnarray}
\delta \rho \left( B_z \right) &=& - \frac{e^2\rho^2 D}{\pi \hbar}
\!\!\! \sum_{s,l=0,x,y,z} \!\!\! c_s c_l \, \frac{eB_z}{\pi\hbar} \sum_{n=0}^{\tau_{B}/\tau}
C_{s}^{l} (q_n) , \label{f2} \\
\Delta \rho \left( B_z \right) &=& \frac{e^2\rho^2}{2\pi h}
\sum_{s,l=0,x,y,z} c_s c_l \,
F \!\left( \frac{B_z}{{\cal B}_{\varphi} + {\cal B}_{s}^{l}} \right) , \label{monrB} \\
{\cal B}_{\varphi} &=& \frac{\hbar c}{4De} \tau_{\varphi}^{-1} \, , \qquad
{\cal B}_{s}^{l} = \frac{\hbar c}{4De} \Gamma_{s}^{l} \, , \nonumber \\
F \!\left( z \right) &=& \ln z + \psi \left( \frac{1}{2} + \frac{1}{z} \right) \, , \nonumber
\end{eqnarray}
where $\psi$ is the digamma function.
In Eq.~(\ref{f2}) the influence of an external field $B_z$ is taken into
account through discrete values $q_n^2 = (n + 1/2)/(D\tau_B)$ where $\tau_B = \hbar/(4DeB_z)$.
Used in conjunction with the Cooperon gaps $\Gamma_{s}^{l}$ listed in Table~\ref{tab:2},
Eqs.~(\ref{f1}-\ref{monrB})
provide a general description of the weak-localization correction and corresponding
low-field magnetoresistance in the presence of SO coupling in graphene,
parameterized by six SO scattering rates
$\tau_{KM}^{-1}$, $\tau_{BR}^{-1}$, $\tau_{zv,\mathrm{e}}^{-1}$,
$\tau_{iv,\mathrm{e}}^{-1}$, $\tau_{zv,\mathrm{o}}^{-1}$, $\tau_{iv,\mathrm{o}}^{-1}$.

In order to analyze the influence of SO coupling in a realistic experimental
situation \cite{tik08,koz11,lara11}, we consider - in the rest of this paper - the
spin-independent intervalley scattering rate $\tau_{iv}^{-1}$ to exceed the
decoherence rate $\tau_{\varphi}^{-1}$ and the rates due to SO coupling.
This means that $\{ \Gamma_{j}^{x} , \Gamma_{j}^{y} , \Gamma_{j}^{z} \} \gg \Gamma_{j}^{0}$
and, thus, the valley-triplet Cooperons in Eqs.~(\ref{monr0}-\ref{monrB}) may be neglected.
Then, the six SO rates may be combined into just two relevant combinations:
a rate $\tau_{\mathrm{sym}}^{-1}$ due to $z \rightarrow - z$ symmetric
SO coupling (terms ${\hat{V}}_{\mathrm{sym}}$ and ${\hat{h}}_{KM}$) and a rate
$\tau_{\mathrm{asy}}^{-1}$ due to $z \rightarrow - z$ asymmetric
coupling (${\hat{V}}_{\mathrm{asy}}$ and ${\hat{h}}_{BR}$):
\begin{eqnarray}
\tau_{\mathrm{sym}}^{-1} &=& \tau_{KM}^{-1} + 2 \tau_{zv,\mathrm{e}}^{-1} + 4 \tau_{iv,\mathrm{e}}^{-1} , \\
\tau_{\mathrm{asy}}^{-1} &=& \tau_{BR}^{-1} + 2 \tau_{zv,\mathrm{o}}^{-1} + 4 \tau_{iv,\mathrm{o}}^{-1} .
\end{eqnarray}
Here, $\tau_{BR}^{-1}$ accounts for the Dyakonov-Perel \cite{dpspin} spin
relaxation contribution, and the other terms for Elliott-Yafet \cite{eyspin} spin relaxation.

The application of an in-plane magnetic field produces an interplay
between SO coupling and Zeeman splitting, as in semiconductor
quantum dots \cite{aleiner01,zumbuhl02}. In-plane magnetic field
${\vec{\ell}} B_{\parallel}$ introduces an additional term in the Hamiltonian
$\delta {\hat H} = (\hbar\omega/2) {\vec{\ell}} {\vec s}$ where
$\omega = 2 \mu_B B_{\parallel}/\hbar$, $\epsilon_z = \hbar \omega$
is the Zeeman energy and ${\vec{\ell}} = (\ell_x , \ell_y , 0)$, $|{\vec{\ell}}|=1$.
This couples the spin-singlet $C_{0}^{0}$ to the triplets $C_{x}^{0}$ and $C_{y}^{0}$.
The spin part of the matrix
equation for the valley singlet Cooperons
$(C_{0}^{0},C_{x}^{0},C_{y}^{0},C_{z}^{0})\equiv \mathbf{C}^{0}$
has the form
\begin{gather*}
\left(
\begin{array}{cccc}
\Pi & -i \omega \ell_{x} & -i \omega \ell_{y} & 0 \\
-i \omega \ell_{x} & \Pi+\tau_{\mathrm{so}}^{-1} & 0 & 0 \\
-i \omega \ell_{y} & 0 & \Pi+\tau_{\mathrm{so}}^{-1} & 0 \\
0 & 0 & 0 & \Pi+2\tau_{\mathrm{asy}}^{-1}
\end{array}%
\right) \mathbf{C}^{0}=1 , \\
\Pi = D \left( i\mathbf{\nabla} + 2e\mathbf{A}/c\hbar \right)^{2} + \tau_{\varphi }^{-1} , \\
\!\!\!\!\!\!\!\!\!\!\!\! \!\!\!\!\!\!\!\!\!\!\!\! \!\!\!\!\!\!\!\!\!\!\!\!
\tau_{\mathrm{so}}^{-1} = \tau_{\mathrm{sym}}^{-1} + \tau_{\mathrm{asy}}^{-1} .
\end{gather*}
After matrix inversion,
\begin{eqnarray*}
C_{0}^{0} \!&=&\! \frac{\Pi+ \tau_{\mathrm{so}}^{-1}}
{\Pi(\Pi + \tau_{\mathrm{so}}^{-1}) + \omega^2}
\xrightarrow[\epsilon_z \to \infty]{} 0  , \\
C_{x/y}^{0} &=& \frac{\Pi(\Pi + \tau_{\mathrm{so}}^{-1})+\omega^2(1-\ell_{x/y}^2)}
{(\Pi + \tau_{\mathrm{so}}^{-1})[\Pi(\Pi + \tau_{\mathrm{so}}^{-1}) + \omega^2]}
\xrightarrow[\epsilon_z \to \infty]{}
\frac{1-\ell_{x/y}^2}
{\Pi + \tau_{\mathrm{so}}^{-1}} , \\
C_{z}^{0} \!&=&\! \frac{1}{\Pi + 2\tau_{\mathrm{asy}}^{-1}}
\xrightarrow[\epsilon_z \to \infty]{} \frac{1}{\Pi + 2\tau_{\mathrm{asy}}^{-1}} ,
\end{eqnarray*}
where the limit $\epsilon_z \to \infty$ of large Zeeman energy essentially means that
$\epsilon_z \gg \hbar \tau_{\mathrm{so}}^{-1}$.

\begin{figure}[t]
\centerline{\epsfxsize=1.0\hsize \epsffile{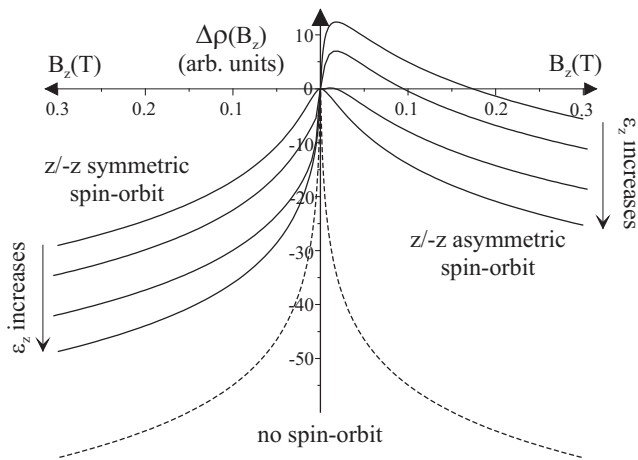}}
\caption{The low-field magnetoresistivity
in the presence of $z \rightarrow -z$ symmetric (left) or asymmetric
(right) SO scattering, as compared to the absence
of SO scattering (lower dashed curves).
Solid curves show the influence of SO scattering,
$\tau_{\mathrm{sym}}^{-1} = 25\tau_{\varphi}^{-1}$ and
$\tau_{\mathrm{asy}}^{-1} = 25\tau_{\varphi}^{-1}$, respectively,
with Zeeman energy $\epsilon_z = 0$ (top)
to $\epsilon_z \gg \tau_{\mathrm{so}}^{-1}$ (bottom).
}
\label{fig:2}
\end{figure}

In the absence of in-plane field, $\epsilon_z = 0$,
the low-field magnetoresistance Eqs.~(\ref{f2},\ref{monrB}) is given by
\begin{eqnarray}
&& \!\!\!\!\!\!\!\!\!
\Delta \rho \!= \!\frac{e^{2}\rho ^{2}}{2\pi h} \!\!
\left[ F\!\! \left( \!\frac{B_z}{{\cal B}_{\varphi}}\! \right)\!
- F\!\! \left( \!\frac{B_z}{{\cal B}_{\varphi }+ {\cal B}_{\mathrm{asy}}}\! \right)\!
-2 F\!\! \left( \!\frac{B_z}{{\cal B}_{\varphi}+{\cal B}_{\mathrm{so}}}\! \right)\! \right] \!\! ,  \nonumber \\
&& {\cal B}_{\mathrm{asy}} = \frac{\hbar c}{2De} \tau_{\mathrm{asy}}^{-1}, \qquad
{\cal B}_{\mathrm{so}} = \frac{\hbar c}{4De} \tau_{\mathrm{so}}^{-1} . \label{result}
\end{eqnarray}
In the absence of SO coupling, ${\cal B}_{\mathrm{so}} = {\cal B}_{\mathrm{asy}} = 0$,
Eq.~(\ref{result}) would describe negative magnetoresistance corresponding
to weak localization \cite{WLmono,WLreview}
(lower dashed curves in Fig.~\ref{fig:2}).
In the presence of $z \rightarrow - z$ symmetric SO coupling only,
${\cal B}_{\mathrm{asy}} = 0$, the contribution
of the third term in Eq.~(\ref{result}) is diminished,
and the first and second terms cancel each other, leading to a suppression of
magnetoresistance for $B_z \alt {\cal B}_{\mathrm{so}}$
(upper solid curve on the left of Fig.~\ref{fig:2})
which mimicks the effect of a saturated value of $\tau_{\varphi}^{-1}$:
$\tau_{\varphi}^{-1} \rightarrow \tau_{\varphi}^{-1} + \tau_{\mathrm{sym}}^{-1}$.
When $z \rightarrow - z$ symmetry is broken,
${\cal B}_{\mathrm{asy}} \neq 0$ and ${\cal B}_{\mathrm{so}} \neq 0$, there is
relaxation of all spin-triplets, and the second and third terms
in Eq.~(\ref{result}) are suppressed, leaving the first (singlet) term
to determine anti-localization behavior at $T \to 0$ with positive magnetoresistance
at low fields $B_z \alt {\cal B}_{\mathrm{asy}}$.

In the limit $\epsilon_z \gg \hbar \tau_{\mathrm{so}}^{-1}$,
\begin{eqnarray}
\Delta \rho \!= \!- \frac{e^{2}\rho ^{2}}{2\pi h} \!\!
\left[
F\!\! \left( \!\frac{B_z}{{\cal B}_{\varphi }+ {\cal B}_{\mathrm{asy}}}\! \right)\!
+ F\!\! \left( \!\frac{B_z}{{\cal B}_{\varphi}+{\cal B}_{\mathrm{so}}}\! \right)\! \right] \! . \label{result2}
\end{eqnarray}
This result shows that for $z \rightarrow - z$ symmetric SO coupling (${\cal B}_{\mathrm{asy}}=0$),
in-plane field partially restores weak localization at the lowest
temperatures lifting the limitation of $\tau_{\varphi}$ discussed above.
In contrast, for $z \rightarrow - z$ asymmetric SO coupling, in-plane field
changes weak anti-localization into a suppressed weak localization behavior.
The low-field magnetoresistance calculated using Eq.~(\ref{f2})
for intermediate values of $\epsilon_z$ is plotted in Fig.~\ref{fig:2},
for $z \rightarrow -z$ symmetric (left) and asymmetric (right) SO scattering.

To summarize, among the two extremes of SO coupling in graphene, $z \rightarrow - z$
symmetric and $z \rightarrow - z$ asymmetric, the manifestation of the latter in quantum transport
resembles that observed in a 2D electron gas in GaAs/AlGaAs heterostructures,
whereas the former is peculiar for graphene. Experimentally, the effect
of $z \rightarrow -z$ symmetric SO coupling can be taken for a decoherence
time ``saturation'' ($\tau_{\varphi} (T \to 0) \rightarrow \tau_{\mathrm{sym}}$)
at low temperatures. Unlike inelastic decoherence, such saturation can be partially lifted
by electron Zeeman splitting induced by a strong in-plane magnetic field making
the negative magnetoresistance $\Delta \rho \left( B_z \right)$ sharper
when $\tau_{\varphi}^{-1} (T \to 0) \rightarrow 0$.
It is necessary to mention that a similar behavior of weak localization magnetoresistance
should be expected in magnetically contaminated conductors \cite{pierre03}.
Spin-flip scattering of electrons from localized spins leads to saturation
of $\tau_{\varphi}$ at the value of the spin-relaxation time whereas in-plane field
freezes local moments thus suppressing spin-flip scattering of electrons and
restoring the full size of the weak localization effect. However, the size of the
in-plane field lifting the ``saturation'' of $\tau_{\varphi}$ in these two
cases is different: polarization of magnetic impurities requires
$\mu_B B_{\parallel} > k T$ whereas the suppression of the effect of
$z \rightarrow - z$ symmetric SO coupling occurs when
$\mu_B B_{\parallel} > \hbar \tau_{\mathrm{sym}}^{-1}$.

This project has been funded by JST-EPSRC Japan-UK Cooperative Programme Grant EP/H025804/1,
EU STREP ConceptGraphene, and the Royal Society.

\end{document}